\documentclass[journal]{IEEEtran}
 
 \usepackage[T1]{fontenc}
 
 \usepackage{amsmath, amsthm, amssymb, amsfonts, bm, mathtools, dsfont, nccmath}
 \usepackage{graphicx, epsfig, epstopdf}
 \usepackage{caption, subcaption}
 \usepackage{float, stfloats}
 
 \usepackage{algpseudocode}
 \usepackage{algorithm}
 
 \usepackage{tabularx, enumerate, xspace, enumitem}
 
 \usepackage[sort,compress]{cite}  
 \usepackage[hidelinks]{hyperref}  
 
 \theoremstyle{plain}

 \usepackage{balance}
 
 \usepackage{xcolor}

 \usepackage[framemethod=tikz]{mdframed}
 \definecolor{mycolor}{rgb}{0.122, 0.435, 0.698}
 \newmdenv[innerlinewidth=0.5pt, roundcorner=4pt, linecolor=mycolor,
 innerleftmargin=6pt, innerrightmargin=6pt,
 innertopmargin=6pt, innerbottommargin=6pt]{mybox}
 
 \newcommand{\bmk}{\mathbf{k}}  
   
   \newcommand{\bmV}{\mathbf{V}}

   \newcommand{\bmW}{\mathbf{W}}

 \newcommand{\bmq}{\mathbf{q}}  \newcommand{\bmr}{\mathbf{r}}

 \usepackage[normalem]{ulem}
 \usepackage[user]{zref}
 
 \newcounter{revc}
 \makeatletter
 \zref@newprop{revcontent}{} \zref@addprop{main}{revcontent}
 \zref@newprop{revsec}{} \zref@addprop{main}{revsec}
 \zref@newprop{revpage}{} \zref@addprop{main}{revpage}
 
 \newcommand{\revi}[2]{%
 	\zref@setcurrent{revsec}{\thesection}%
 	\zref@setcurrent{revpage}{\thepage}%
 	\zref@setcurrent{revcontent}{#2}%
 	\refstepcounter{revc}%
 	\label{#1}%
 	\zlabel{#1}%
 	\textcolor{blue}{#2}%
 }
 
 \newcommand{\revinu}[2]{%
 	\zref@setcurrent{revsec}{\thesection}%
 	\zref@setcurrent{revcontent}{#2}%
 	\refstepcounter{revc}%
 	\zlabel{#1}%
 	\label{#1}
 	#2 }
 
 \newcommand{\revr}[2]{%
 	\zref@setcurrent{revsec}{\thesection}%
 	\zref@setcurrent{revcontent}{#2}%
 	\refstepcounter{revc}%
 	\zlabel{#1}%
 	\label{#1} \sout{#2}} 
 \makeatother

 \expandafter\def\expandafter\quote\expandafter{\quote\onehalfspacing\fontsize{12}{14}\selectfont}
 
 \def\BibTeX{{\rm B\kern-.05em{\sc i\kern-.025em b}\kern-.08em
 		T\kern-.1667em\lower.7ex\hbox{E}\kern-.125emX}}
 
 \DeclareUnicodeCharacter{2212}{\textendash}
 \DeclareUnicodeCharacter{03A8}{\textendash}

\begin{document}

\title{Joint Beamforming and 3D Location Optimization for Multi-User Holographic UAV Communications}  
\author{Chandan~Kumar~Sheemar,~\IEEEmembership{Member,~IEEE}, Asad Mahmood,~\IEEEmembership{Member,~IEEE},  \\Christo Kurisummoottil Thomas,~\IEEEmembership{Member,~IEEE}, George C. Alexandropoulos,~\IEEEmembership{Senior Member,~IEEE}, \\Jorge Querol,~\IEEEmembership{Member,~IEEE}, Symeon Chatzinotas,~\IEEEmembership{Fellow,~IEEE}, Walid Saad,~\IEEEmembership{Fellow,~IEEE}  
	\thanks{Chandan Kumar Sheemar, Asad Mahmood, and Symeon Chatzinotas are with the SnT department at the University of Luxembourg (email: \{chandankumar.sheemar, asad.mahmood, symeon.chatzinotas\}@uni.lu). George C. Alexandropoulos is with the Department of Informatics and Telecommunications, National and Kapodistrian University of Athens, 15784 Athens, Greece (email: alexandg@di.uoa.gr). Christo Kurisummoottil Thomas and Walid Saad are with Virginia Tech (emails: \{christokt,walids\}@vt.edu). } 
 
} 
 \maketitle

\begin{abstract}
This paper pioneers the field of multi-user holographic unmanned aerial vehicle (UAV) communications, laying a solid foundation for future innovations in next-generation aerial wireless networks. The study focuses on the challenging problem of jointly optimizing hybrid holographic beamforming and 3D UAV positioning in scenarios where the UAV is equipped with a reconfigurable holographic surface (RHS) instead of conventional phased array antennas. Using the unique capabilities of RHSs, the system dynamically adjusts both the position of the UAV and its hybrid beamforming properties to maximize the sum rate of the network. To address this complex optimization problem, we propose an iterative algorithm combining zero-forcing digital beamforming and a gradient ascent approach for the holographic patterns and the 3D position optimization, while ensuring practical feasibility constraints. The algorithm is designed to effectively balance the trade-offs between power, beamforming, and UAV trajectory constraints, enabling adaptive and efficient communications, while assuring a monotonic increase in the sum-rate performance. Our numerical investigations demonstrate that the significant performance improvements with the proposed approach over the benchmark methods, showcasing enhanced sum rate and system adaptability under varying conditions. 
\end{abstract}
\begin{IEEEkeywords}
Holographic beamforming, holographic UAVs, hybrid beamforming, sum-rate maximization. 
\end{IEEEkeywords}

\IEEEpeerreviewmaketitle

\section{Introduction} \label{Intro} 
\IEEEPARstart{U}{nmanned} Aerial Vehicles (UAVs), commonly known as drones, are emerging as a pivotal component of 6G wireless communication networks, addressing the need for enhanced connectivity and adaptability in diverse environments \cite{mozaffari2019tutorial}. As 6G aspires to deliver ultra-reliable, low-latency communication, high data rates, and massive connectivity, UAVs offer unique advantages due to their mobility, flexibility, and ability to provide on-demand network services \cite{mozaffari2018beyond}. UAVs can function as airborne base stations, relays, or mobile users, effectively extending network coverage to remote or underserved areas and alleviating congestion in densely populated regions \cite{yi20233}. Their ability to establish line-of-sight (LoS) communications with ground users and other network nodes ensures high-quality links and improved performance even in challenging wireless environments~\cite{wu2018joint}.

However, the deployment of UAVs in 6G networks also faces significant challenges. Energy consumption remains a critical concern, as UAVs rely on limited onboard battery capacity, restricting their operational time and coverage range \cite{zeng2017energy}. Prolonged missions demand efficient energy management strategies, such as optimizing flight paths, reducing hovering times \cite{mozaffari2017wireless}, or incorporating solar panels and wireless charging stations. Moreover, ensuring secure and reliable communication links is a challenge, especially in environments with high interference or physical obstructions \cite{azari2022thz}. UAVs also face constraints in payload capacity, limiting their ability to carry advanced communication equipment or additional sensors.

In parallel, reconfigurable holographic surfaces (RHS) have emerged
as a promising solution for next-generation wireless systems \cite{gong2023holographic,iacovelli2024holographic,wei2023tri,zhang2022holographic,AXN2023,XYA2024}. They can operate as dual-function arrays capable of transmitting and receiving electromagnetic waves, seamlessly integrated with the transceiver to facilitate beam steering in desired directions \cite{huang2020holographic,zhang2022holographic}. The compact structure of RHS, fabricated using printed circuit board (PCB) technology, incorporates a feed mechanism that generates an electromagnetic reference wave. This reference wave propagates across the metasurface and radiates energy through individual radiation elements. The metasurface employs engineered metamaterials to form a holographic pattern, leveraging holographic interference between the reference wave and a target object wave \cite{gong2023holographic,deng2022holographic,HMIMO_computing}. By modulating this holographic pattern, each element adjusts the radiation weighting of the reference wave to synthesize highly directional beams. This technique, known as holographic beamforming, eliminates the need for conventional phased array antenna (PAAs) systems \cite{sheemar2022practical}, which rely on complex phase-shifting circuits and bulky mechanical components \cite{yetis2021joint,rosson2019towards}.

The RHS-assisted transceivers can offer several notable advantages over traditional PAAs based designs \cite{black2017holographic,shlezinger2021dynamic}. First, their compact design based on PCB technology significantly reduces both the size and weight compared to phased arrays \cite{lin2019hybrid} and dish antennas, making it easier to integrate into transceiver systems. Second, unlike conventional systems, RHSs do not require active amplification or complex phase-shifting circuits \cite{payami2016hybrid}, which results in lower power consumption. Additionally, since the components used in the RHS (e.g., varactor diodes, PCBs, and DC control circuits) are readily available off-the-shelf, the overall production cost is considerably lower. These features make the RHS a highly cost-effective and energy-efficient solution, positioning it as a promising alternative to traditional PAAs \cite{deng2022hdma,ghermezcheshmeh2023parametric}. Consequently, RHS-assisted UAVs can become a cornerstone for next-generation aerial communications with greater performance, energy efficiency, and enhanced hovering times. 

\subsection{State-of-the-Art}
While RHS-based systems present remarkable advantages \cite{di2021reconfigurable}, their development remains in its nascent stages.
In \cite{deng2021reconfigurable}, the problem of hybrid holographic beamforming for sum-rate maximization in an RHS-based system is tackled. The study introduces a solution by jointly optimizing the digital beamformer and the analog holographic beamformer. For the latter, the non-convex optimization problem is approached by breaking it down into a series of convex subproblems, which are iteratively solved. In \cite{hu2023holographic}, an alternative solution is proposed that employs fully analog holographic beamforming to maximize the sum rate without digital beamforming. This approach eliminates the need for expensive radio frequency (RF) chains, reducing cost and power consumption and making it particularly appealing for energy-efficient applications. However, such a solution offers less beam steering and interference management flexibility. Further extending the understanding of RHS systems, \cite{an2023tutorial} delves into the spatial degrees of freedom (DoF) and ergodic capacity of a point-to-point RHS setup. By analyzing the fundamental capacity limits of RHS, this work provides important theoretical benchmarks that can guide the design of future systems. In addition to these works, \cite{you2022energy} explores the energy-efficiency maximization problem in RHS systems, focusing on the case where only analog holographic beamforming is employed. The study proposes solutions under both perfect and imperfect channel state information (CSI) scenarios, emphasizing the practical importance of robust and efficient system designs in real-world settings. In \cite{shlezinger2021dynamic}, the authors conducted an in-depth analysis of the operational principles associated with dynamic metasurface antennas (DMAs), which represent a metasurface-based implementation framework for HMIMO transceivers. These studies further explored the advantages and capabilities of DMAs in the context of 6G communications, highlighting their potential and identifying key challenges arising from their broad range of prospective applications. The study in \cite{zhu2024electromagnetic} explored electromagnetic (EM) information theory inspired by HMIMO systems, presenting foundational principles of this interdisciplinary framework, including modeling, analytical methods, and practical applications. In \cite{wan2021terahertz}, the potential of HMIMO for terahertz (THz) communications is investigated. 

Further studies exploring the potential of RHS-based holographic systems for near-field communications have demonstrated significant advancements in this area \cite{XYA2024,ji2023extra,gan2022near,gong2024holographic,wei2023tri,GIS2023,GA2024,GA2024a}. In \cite{ji2023extra}, the focus is on spatially constrained antenna apertures with rectangular symmetry, aiming to enhance spatial degrees of freedom and channel capacity. The study achieves notable performance improvements by leveraging evanescent waves for information transmission in near-field scenarios and utilizing Fourier plane-wave series expansion. In \cite{gan2022near}, the integration of RHS into millimeter-wave systems is investigated for advanced beamforming and localization. This work capitalizes on near-field wavefront manipulation to optimize beamforming patterns and reduce localization errors, thereby improving communication quality and positioning accuracy. A generalized electromagnetic-domain near-field channel model is introduced in \cite{gong2024holographic}, where the authors analyze capacity limits for point-to-point holographic systems with arbitrarily placed surfaces in a line-of-sight (LOS) environment. The study presents two efficient channel models—one offering high precision with a sophisticated formulation and the other emphasizing computational simplicity with minor accuracy trade-offs. Additionally, a tight upper bound on system capacity is derived using an analytical framework. Lastly, \cite{wei2023tri} explores the use of triple polarization (TP) in multi-user wireless communication systems employing RHS to enhance capacity and diversity without increasing antenna array size. A TP near-field channel model based on the dyadic Green's function is developed, and two precoding schemes are proposed to mitigate cross-polarization and inter-user interference, offering innovative solutions for improving near-field multi-user communication efficiency.

The only study to consider holographic UAV communication is available in \cite{song2024miniature}. The authors considered the case of one
source node communicating with two terminals:
an RHS-assisted UAV and a receiver destination node. The problem of maximizing the energy efficiency is considered for which a novel optimization method is proposed.

\subsection{Motivation and Contributions}
The integration of RHS into UAV communications represents a transformative opportunity that remains largely unexplored in the existing literature. UAVs face stringent constraints on energy consumption, payload weight, and flight duration. The inherent advantages of RHSs, including exceptional energy efficiency and significantly reduced transceiver weight, directly address these challenges. RHSs enable highly flexible and efficient beamforming with minimal hardware complexity, eliminating the need for power-intensive RF components. This reduction in power demand can not only conserve the UAV's limited battery capacity but also extend operational endurance. Furthermore, the lightweight design of RHSs can minimize the overall payload, improving flight dynamics and reducing propulsion energy requirements. By combining energy efficiency with lightweight architecture, RHSs can significantly enhance the capabilities of UAVs, enabling prolonged flight durations, expanded coverage areas, and more sustainable communication networks. These attributes make RHS-assisted UAV systems a compelling solution for next-generation wireless networks, paving the way for innovative applications in areas such as remote connectivity, disaster response~\cite{YKK2024}, and beyond.

This work introduces a comprehensive framework designed to enhance the performance of UAV-enabled wireless communication systems by integrating RHSs at the analog front-end. The UAV operates within a three-dimensional (3D) space, serving multiple ground users, and employs a hybrid transceiver architecture that combines low-dimensional digital beamforming with large-dimensional holographic beamforming facilitated by an RHS. The core objective of this study is the joint optimization of the UAV's 3D position, holographic beamforming weights, and digital beamformers to maximize the system's sum rate. The optimization framework is subject to critical constraints, including the UAV's operational area, altitude limits, total transmit power, and the realistic weight restrictions of the holographic beamforming. To address this complex optimization problem, a novel alternating optimization framework is proposed, decomposing the task into three interdependent subproblems that are solved iteratively. Unlike prior work, which primarily focuses on UAV's movement optimization based on reconfigurable surface approximations with linear arrays—relying solely on the angle of departure (AoD) \cite{li2023joint}—this study accounts for the practical uniform planar array (UPA) response of RHS. This approach necessitates precise optimization of the UAV's movement, considering the variations in both elevation and azimuth angles relative to each user, together with the path loss. An exact gradient for the 3D position optimization is derived, capturing these angular variations combined with the variation in the path loss, and thus enabling exact UAV movement updates in the optimal direction that maximizes the holographic UAV network's sum rate. Finally, extensive simulation results validate the proposed framework's effectiveness and robustness across diverse system configurations and operational scenarios.

\emph{Paper Organization:} The structure of this paper is as follows: Section \ref{section_2} presents the system model and formulates the optimization problem for a joint holographic beamforming and 3D location optimization of an RHS-assisted UAV. Section \ref{section_3} details the proposed algorithmic framework for addressing the joint optimization problem and provides the complexity analysis. Section \ref{section_4} provides a comprehensive discussion of the simulation results, while Section \ref{section_5} presents the final conclusions of the paper.

\emph{Notations:} Scalars are denoted by lowercase or uppercase letters, while vectors and matrices are represented by bold lowercase and bold uppercase letters, respectively. The transpose, Hermitian transpose, and inverse of a matrix $\mathbf{A}$ are denoted by $\mathbf{A}^\mathrm{T}$, $\mathbf{A}^\mathrm{H}$, and $\mathbf{A}^{-1}$, respectively. Sets are indicated by calligraphic letters (e.g., $\mathcal{A}$), and their cardinality is represented by $|\mathcal{A}|$. Finally, $\|\cdot\|$ denotes the $l_2$-norm.  

\section{System and Channel Models} \label{section_2}
\begin{figure} 
    \centering
    \includegraphics[width=\linewidth]{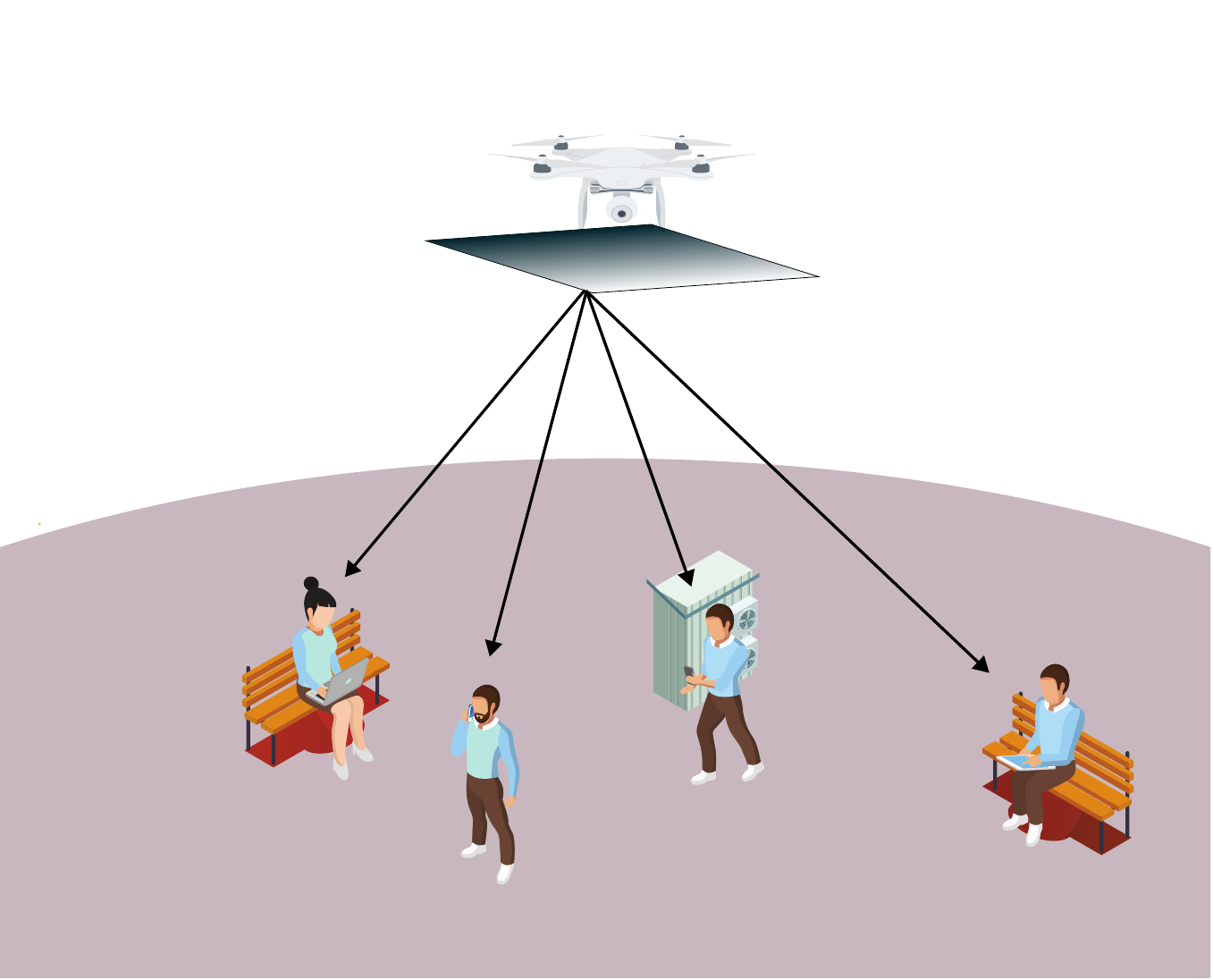}
    \caption{The proposed multi-user holographic UAV communications realizing hybrid beamforming.}
    \label{rate_64elements}
\end{figure}

We consider a UAV-enabled wireless communication system consisting of $ D $ ground users and a UAV equipped with an RHS for communication at the antenna front end. The ground users are located on the $(x,y)$ plane at positions $ \mathbf{u}_d = [u_{d,x}, u_{d,y}]^T \in \mathbb{R}^2 $, where $ d = 1, \dots, D $, while the UAV 3D location is parameterized in the three-dimensional coordinates as $ \mathbf{q} = [q_x, q_y, q_z]^T \in \mathbb{R}^3 $. The RHS mounted on the UAV comprises $ M = \sqrt{M} \times \sqrt{M} $ reconfigurable elements, where $ \sqrt{M} $ denote the number of reconfigurable elements along the $x$-axis and y-axis, respectively. The UAV employs a hybrid beamforming architecture combining digital and analog beamforming to serve ground users efficiently. 

We assume that the number of feeds equals the number of RF chains. In the baseband, the UAV implements digital beamforming, and the beamformer for user $ d $ is denoted as $ \mathbf{v}_d \in \mathbb{C}^{K \times 1} $, with $ K $ representing the number of RF chains available on the UAV. The number of RF chains in the hybrid RHS architecture is assumed to satisfy $ K \geq D $, ensuring that the system can support active data streams. Each RF chain feeds its corresponding RHS feed, transforming the high-frequency electrical signal into a propagating electromagnetic wave. The radiation amplitude of the reference wave in each metamaterial element of the RHS is dynamically controlled using a holographic beamforming matrix $ \mathbf{W} \in \mathbb{C}^{M \times K} $, enabling the generation of the desired directional beams. Here, $ M $ denotes the total number of elements of the holographic surface, with each element represented as \cite{deng2021reconfigurable}:
\begin{equation}
  W_{m} =  w_{m} \cdot e^{-j \bm{k}_s \bm{r}_{m}^k}  
\end{equation}
where $ e^{-j \bm{k}_s \cdot \bm{r}_{m}^k} $ corresponds to the phase of the reference wave, and $ \bm{k}_s $ and $ \bm{r}_{m,n}^k $ represent the wave vector and position vector of the corresponding metamaterial element, respectively.

The received signal at user $d$ given the position of the UAV $\bmq$ can be written as
\begin{equation}
    y_d = \mathbf{h}_d(\mathbf{q})^H \mathbf{W} \mathbf{v}_d s_d 
+ \sum_{k \neq d} \mathbf{h}_d(\mathbf{q})^H \mathbf{W} \mathbf{v}_k s_k + n_d.
\end{equation}

 The term $ \mathbf{v}_d \in \mathbb{C}^K $ denotes the digital beamforming vector for user $ d $, $ s_d \sim \mathcal{CN}(0, 1) $ is the transmitted symbol for user $ d $, and $ n_d \sim \mathcal{CN}(0, \sigma^2) $ is the additive noise.

\subsection{Channel Model}

 The channel vector $ \mathbf{h}_d(\bmq) $ between the UAV and user $ d $ is modelled as
\begin{equation}
\mathbf{h}_d = \sqrt{\frac{\beta_0}{\|\mathbf{q} - \mathbf{u}_d\|^2}} \cdot \mathbf{a}(\theta^d, \phi^d),
\end{equation}
where $ \beta_0 $ is the reference channel gain, and $ \|\mathbf{q} - \mathbf{u}_d\| $ represents the distance between the UAV and user $ d $. The term $ \mathbf{a}(\theta, \phi) $ is the UPA response vector of the RHS, which is defined as:
\begin{equation}
\mathbf{a}(\theta^d, \phi^d) = \mathbf{a}_y(k_y^d) \otimes \mathbf{a}_x(k_x^d),
\end{equation}
where $ \otimes $ denotes the Kronecker product, and $ \mathbf{a}_x(k_x^d) $ and $ \mathbf{a}_y(k_y^d) $ represent the array response vectors along the $ x $- and $ y $-axes, respectively, and given as

\begin{equation}
    \mathbf{a}_x(\theta^d, \varphi^d) = 
    \begin{bmatrix}
    1 \\
    e^{j k_f d_x \sin\theta^d \cos\varphi^d} \\
    e^{j 2 k_f d_x \sin\theta^d \cos\varphi^d} \\
    \vdots \\
    e^{j (\sqrt{M}-1) k_f d_x \sin\theta^d \cos\varphi^d}
    \end{bmatrix},
\end{equation}
\begin{equation}
    \mathbf{a}_y(\theta^d, \varphi^d) = 
    \begin{bmatrix}
    1 \\
    e^{j k_f d_y \sin\theta^d \sin\varphi^d} \\
    e^{j 2 k_f d_y \sin\theta^d \sin\varphi^d} \\
    \vdots \\
    e^{j (\sqrt{M}-1) k_f d_y \sin\theta^d \sin\varphi^d}
    \end{bmatrix}.
\end{equation}
where $ k_f$ dentoes the wavenumber is free space, and $ d_x $ and $ d_y $ denote the spacing between adjacent elements along the $ x $- and $ y $-axes, respectively, and $ \sin\theta^d \cos\varphi^d $ and $ \sin\theta^d \sin\varphi^d $ represent the directional cosines corresponding to the azimuth and elevation angles. These angles can be determined geometrically, given the position of the UAV and user $d$ as
\begin{subequations}
\begin{equation}
    \theta^d = \arctan\left(\frac{q_z}{\sqrt{(q_x - u_{d,x})^2 + (q_y - u_{d,y})^2}}\right),
\end{equation}
\begin{equation}
    \phi^d = \arctan2\left(q_y - u_{d,y}, q_x - u_{d,x}\right).
\end{equation}
\end{subequations}

\subsection{Problem Formulation}

The signal-to-interference-plus-noise ratio (SINR) for user $ d $ is defined as  
\[
\text{SINR}_d = \frac{\left| \mathbf{h}_d^H(\mathbf{q}) \mathbf{W} \mathbf{v}_d \right|^2}{\sum_{k \neq d} \left| \mathbf{h}_d^H(\mathbf{q}) \mathbf{W} \mathbf{v}_k \right|^2 + \sigma^2},
\]
where $ \mathbf{h}_d(\mathbf{q})^H $ represents the channel vector between the UAV at position $ \mathbf{q} $ and user $ d $, $ \mathbf{W} $ is the holographic beamforming matrix, $ \mathbf{v}_d $ is the digital beamforming vector for user $ d $, and $ \sigma^2 $ denotes the noise power. 

 We aim to maximize the system's sum rate subject to the total power constraint for the digital beamformers and the real-valued amplitude constraint on the holographic beamformer. Furthermore, we consider that the RHS-assisted UAV is capable of maneuvering within a designated area $\mathcal{A}$ on the $(x, y)$-plane, while its altitude is restricted within a predefined range $z_{\text{min}} \leq q_z \leq z_{\text{max}}$. Under these constraints, the joint optimization problem can be formulated as follows:

\begin{subequations}
    \begin{equation}
        \max_{\mathbf{q}, \mathbf{W}, \bmV} \quad \sum_{d=1}^D \log_2\left(1 + \frac{\left| \mathbf{h}_d(\mathbf{q})^H \mathbf{W} \mathbf{v}_d \right|^2}{\sum_{k \neq d} \left| \mathbf{h}_d(\mathbf{q})^H \mathbf{W} \mathbf{v}_k \right|^2 + \sigma_d^2}\right),
    \end{equation}
    \begin{equation} \label{c1}
        \text{s.t.} \quad \sum_{d=1}^D \|\mathbf{v}_d\|^2 \leq P_{\text{max}},
    \end{equation}
    \begin{equation} \label{c2}
        0 \leq w_m \leq 1, \quad \forall m,
    \end{equation}
    \begin{equation} \label{c3}
        (q_x, q_y) \in \mathcal{A},
    \end{equation}
    \begin{equation} \label{c4}
        z_{\text{min}} \leq q_z \leq z_{\text{max}}.
    \end{equation}
\end{subequations}

In this formulation, the objective function represents the sum rate of the system. The constraint \eqref{c1} enforces the total power limitation for the digital beamformers, while the constraint \eqref{c2} ensures the real-valued amplitude restrictions for the elements of the holographic beamformer. Constraints \eqref{c3} and \eqref{c4} define the allowable movement region of the UAV on the $(x, y)$-plane and its altitude bounds in 3D space, respectively.

\section{Joint Beamforming and 3D Location Optimization} \label{section_3}
The optimization problem is addressed using an alternating optimization framework, which decomposes the complex, non-convex joint optimization into three interdependent subproblems. This iterative approach simplifies the problem by optimizing one set of variables at a time while keeping the others fixed, ensuring gradual and systematic improvement of the objective function. The three subproblems tackled in each iteration are as follows.

\subsection{Digital Beamforming Optimization}
We assume the holographic beamformer and the position of the UAV fixed.
 To optimize $\bmV = [\mathbf{v}_1, \dots, \mathbf{v}_D]$, we adopt the zero-forcing method. The goal of such a method is to zero-force the interference between users by directing the signal energy toward the intended users while nulling it in directions that could interfere with other users. The effective channel matrix $\mathbf{H}^H(\mathbf{q}) \mathbf{W}$, which combines the UAV-to-user channels and the holographic beamformer, is used to compute the beamforming vectors. Specifically, the digital beamforming matrix for zero-forcing can be optimized as the pseudo-inverse of the effective channel matrix as
\begin{equation}
\mathbf{V} =  \left(\mathbf{H}(\mathbf{q})^H \mathbf{W}\right)^H \left(\mathbf{H}(\mathbf{q})^H \mathbf{W} \left(\mathbf{H}(\mathbf{q})^H \mathbf{W}\right)^H\right)^{-1}.
\end{equation}
 
This solution does not inherently constrain the transmit power. Therefore, a power normalization step is applied to meet the total power constraint with equality as
\begin{equation}
    \mathbf{v}_d = \mathbf{v}_d \sqrt{\frac{P_{\text{max}}}{\sum_{d=1}^D \|\mathbf{v}_d\|^2}}, \quad \forall d \in \{1, \dots, D\}.
\end{equation}

This scaling ensures that the total transmit power does not exceed $P_{\text{max}}$ while preserving the directional properties of the beamforming vectors.

\subsection{ Holographic Beamforming Optimization}

For a fixed UAV position $\mathbf{q}$ and predetermined digital beamforming vectors $\{\mathbf{v}_d\}_{d=1}^D$, the focus shifts to optimizing the system sum rate $R_{\text{sum}}$ by optimizing the holographic weights. The optimization problem for such purpose, assuming the other variables are fixed, can be formulated as follows
 
\begin{subequations}
    \begin{equation}
        \max_{\mathbf{W}} \quad  \sum_{d=1}^D \log_2\left(1 + \frac{\left| \mathbf{h}_d(\mathbf{q})^H \mathbf{W} \mathbf{v}_d \right|^2}{\sum_{k \neq d} \left| \mathbf{h}_d(\mathbf{q})^H \mathbf{W} \mathbf{v}_k \right|^2 + \sigma^2}\right),
    \end{equation}
    \begin{equation}
        \textbf{s.t.} \quad 0 \leq w_m \leq 1, \quad \forall m = 1, 2, \ldots, M.
    \end{equation}
\end{subequations}
Note that the holographic beamformer is structured as $\{w_{m} \cdot e^{-j \bmk_s \cdot \bmr_{m}^k}\}$, where $e^{-j \bmk_s \cdot \bmr_{m}^k}$ represents the fixed phase component of the reference wave originating from the $k$-th feed. Since the phase component is predetermined, the primary objective is to optimize the real-valued holographic weights $w_m$ for each element of the RHS to maximize the sum rate.

In the given holographic architecture, the beamformer can be expressed as a product of two components \cite{zhang2022holographic}:
\begin{equation}
    \bmW = \text{diag}([w_1, w_2, \dots, w_M]) \mathbf{\Phi},
\end{equation}
where:
\begin{itemize}
    \item $\text{diag}([w_1, w_2, \dots, w_M])$ is a diagonal matrix containing the adjustable real-valued weights $w_m$ for each reconfigurable element, with $0 \leq m \leq M$.
    \item $\mathbf{\Phi}$ is a fixed matrix that incorporates the phase components of the reference wave, with elements of the form $e^{-j \bmk_s \cdot \bmr_{m}^k}$, determined by the wave propagation within the RHS waveguide and the distance from feed $k$ to the element $m$.
\end{itemize}

Based on the structure of the holographic beamformer, we first decompose the received signal to highlights its dependence on the holographic weights.

 \subsubsection{Decomposition of the Received Signal}
Recall that the received signal at user $ d $ when the UAV is at position $\bmq$ is given as
\begin{equation}
    y_d = \mathbf{h}_d(\mathbf{q})^H \mathbf{W} \mathbf{v}_d s_d 
+ \sum_{k \neq d} \mathbf{h}_d(\mathbf{q})^H \mathbf{W} \mathbf{v}_k s_k + n_d.
\end{equation}
By expanding the terms for each element $ w_m $, the received signal $ y_d $ can be decomposed as
\begin{equation}
\begin{aligned}
    y_d = \sum_{m=1}^M w_m & h_{d,m}^*(\bmq)(\mathbf{\Phi} \mathbf{v}_d)_m s_d 
\\& + \sum_{k \neq d} \sum_{m=1}^M w_m h_{d,m}^*(\bmq)(\mathbf{\Phi} \mathbf{v}_k)_m s_k + n_d,
\end{aligned}
\end{equation}
where $ h_{d,m}^*(\bmq) $ represents the $ m $-th element of the channel vector $ \mathbf{h}_d(\mathbf{q})^H $, and $ (\mathbf{\Phi} \mathbf{v}_d)_m $ is the $ m $-th element of the product $ \mathbf{\Phi} \mathbf{v}_d $.

Given the decomposed received signal mode linked with each element $w_m$, the SINR for user $ d $ can be written as
\begin{equation}
    \text{SINR}_d = \frac{ \left| \sum_{m=1}^M w_m h_{d,m}^*(\bmq)(\mathbf{\Phi} \mathbf{v}_d)_m \right|^2 }{ \sum_{k \neq d} \left| \sum_{m=1}^M w_m h_{d,m}^*(\bmq)(\mathbf{\Phi} \mathbf{v}_k)_m \right|^2 + \sigma_d^2 }.
\end{equation}

This decomposition provides a clear representation of how the adjustable coefficients $ w_m $ influence the received signal components. Given this decomposition, we can restate the holographic beamforming optimization problem as 

\begin{subequations}
    \begin{equation}
        \max_{w_m} \quad  \sum_{d=1}^D \log_2\left(1 + \frac{ \left| \sum_{m=1}^M w_m h_{d,m}^*(\bmq)(\mathbf{\Phi} \mathbf{v}_d)_m \right|^2 }{ \sum_{k \neq d} \left| \sum_{m=1}^M w_m h_{d,m}^*(\bmq)(\mathbf{\Phi} \mathbf{v}_k)_m \right|^2 + \sigma_d^2 }\right),
    \end{equation}
    \begin{equation}
        \textbf{s.t.} \quad 0 \leq w_m \leq 1, \quad \forall m = 1, 2, \ldots, M.
    \end{equation}
\end{subequations}
To tackle this problem, we employ a gradient-ascent optimization technique to maximize the system's sum rate iteratively. 

\subsubsection{Gradient Derivation for Holographic Beamforming}

To optimize the sum rate of the system assisted by an RHS, we first derive the gradient of the total sum rate $ R_{\text{sum}} $ with respect to the holographic weights $ w_m $. Let $\text{SINR}_d(w_m)$ denote the SINR for user $d$ as a function of the holographic weight $w_m$ and $ N_d(w_m) $ and $D_d(w_m)$ denote its numerator and denominator, respectively.

To optimize the total system sum rate $ R_{\text{sum}} $ with respect to the holographic weight $ w_m $, we derive its gradient by applying the chain rule of differentiation. This process leads to the following expression for the gradient:

\begin{equation}
    \frac{\partial R_{\text{sum}}}{\partial w_m} = \frac{1}{\ln(2)} \sum_{d=1}^D \frac{1}{1 + \text{SINR}_d(w_m)} \frac{\partial \text{SINR}_d(w_m)}{\partial w_m}.
\end{equation}

We further expand $ \frac{\partial \text{SINR}_d(w_m)}{\partial w_m} $ using the chain rule again, which can be computed as

\begin{equation}
    \frac{\partial \text{SINR}_d(w_m)}{\partial w_m} = \frac{D_d(w_m) \frac{\partial N_d(w_m)}{\partial w_m} - N_d(w_m) \frac{\partial D_d(w_m)}{\partial w_m}}{D_d(w_m)^2}.
\end{equation}

A detailed derivation for finding $\frac{\partial \text{SINR}_d(w_m)}{\partial w_m}$is provided in the Appendix \ref{appendix_HBF}, and the final gradient for the sum-rate is given in \eqref{grad_holo}.

\begin{figure*}
\begin{equation}\label{grad_holo}
\begin{aligned}
     \frac{\partial R_{\text{sum}}}{\partial w_m} =  \frac{1}{\ln(2)} \sum_{d=1}^D \frac{1}{1 + \text{SINR}_d(w_m)} \cdot& \frac{\left(h_{d,m}^*(\mathbf{q})(\mathbf{\Phi} \mathbf{v}_d)_m u_d^* + h_{d,m}(\mathbf{q})(\mathbf{\Phi} \mathbf{v}_d)_m^* u_d\right) \left( \sum_{k \neq d} \left| \sum_{m=1}^M w_m h_{d,m}^*(\bmq)(\mathbf{\Phi} \mathbf{v}_k)_m \right|^2 + \sigma^2\right)}{\left(\sum_{k \neq d} \left| \sum_{m=1}^M w_m h_{d,m}^*(\bmq)(\mathbf{\Phi} \mathbf{v}_k)_m \right|^2 + \sigma_d^2\right)^2}  \nonumber  
   \\& - \frac{\left| \sum_{m=1}^M w_m h_{d,m}^*(\bmq)(\mathbf{\Phi} \mathbf{v}_d)_m \right|^2  \sum_{k \neq d} \left(u_{d,k}^* h_{d,m}^*(\mathbf{q})(\mathbf{\Phi} \mathbf{v}_k)_m + u_{d,k} h_{d,m}(\mathbf{q})(\mathbf{\Phi} \mathbf{v}_k)_m^*\right)}{\left(\sum_{k \neq d} \left| \sum_{m=1}^M w_m h_{d,m}^*(\bmq)(\mathbf{\Phi} \mathbf{v}_k)_m \right|^2 + \sigma_d^2\right)^2}.
\end{aligned} 
\end{equation} \hrulefill
\end{figure*}

At each iteration, the weights \(w_m\) are updated by following the gradient of the sum rate with respect to \(w_m\). The gradient represents the steepest ascent direction, ensuring an increase in the sum rate with every update step. To satisfy the physical constraint that \(w_m\) remains within the range \([0, 1]\), the updated value is projected onto the feasible set. The update rule for \(w_m\) at iteration \(t+1\) is given by

\begin{equation}
    w_m^{(t+1)} = \min \left( 1, \max \left( 0, w_m^{(t)} + \alpha \frac{\partial R_{\text{sum}}}{\partial w_m}\bigg|_{\mathbf{w}^{(t)}} \right) \right),
\end{equation}

where \(\eta > 0\) is the learning rate controlling the step size, and \(\frac{\partial R_{\text{sum}}}{\partial w_m}\) is the partial derivative of the sum rate \(R_{\text{sum}}\) with respect to \(w_m\), derived from the system's SINR expressions. The \(\max(0, \cdot)\) operation enforces a lower bound of zero, while the \(\min(1, \cdot)\) operation ensures the weights do not exceed one. 
 Formally, the algorithm for holographic beamforming optimization is provided in Algorithm \ref{holographic_alg}.

\begin{algorithm}
\caption{Proposed Holographic Beamforming}
\begin{enumerate}
    \item \textbf{Initialization}: Choose an initial $\mathbf{w}^{(0)}$ such that $0 < w_m \leq 1$ for all $m = 1, \ldots, M.$
    \item \textbf{Gradient Computation}: At iteration $t$, compute $\frac{\partial R_{\text{sum}}}{\partial w_m}\big|_{\mathbf{w}^{(t)}}$ for all $m$ using the derived expressions.
    \item \textbf{Update Step}:
    \[
    w_m^{(t+1)} = w_m^{(t)} + \eta \frac{\partial R_{\text{sum}}}{\partial w_m}\bigg|_{\mathbf{w}^{(t)}}
    \]
    where $\eta> 0$ is a suitably chosen step size.
    \item \textbf{Projection Step}: If any $w_m^{(t+1)}$ falls outside the feasible range, project it back:
    \[
    w_m^{(t+1)} \leftarrow \min(\max(w_m^{(t+1)},0),1).
    \]
    \item \textbf{Convergence Check}: If $|R_{\text{sum}}(\mathbf{w}^{(t+1)}) - R_{\text{sum}}(\mathbf{w}^{(t)})|$ or $\|\mathbf{w}^{(t+1)} - \mathbf{w}^{(t)}\|$ is below a given threshold, stop. Otherwise, set $t \leftarrow t+1$ and return to Step 2.
\end{enumerate}\label{holographic_alg}
\end{algorithm}

\subsection{UAV 3D Position Optimization with RHS}

In this section, we aim to optimize the 3D position of the UAV, denoted as $ \mathbf{q} = [q_x, q_y, q_z]^T $, to maximize the system's sum rate under practical constraints, given the digital and the holographic beamformers fixed. The optimization problem with respect to $\mathbf{q}$ can be formally stated as
 
\begin{subequations}
    \begin{equation}
        \max_{\mathbf{q}} \quad \sum_{d=1}^D \log_2\left(1 +  \frac{\left| \mathbf{h}_d(\mathbf{q})^H \mathbf{W} \mathbf{v}_d \right|^2}{\sum_{k \neq d} \left| \mathbf{h}_d(\mathbf{q})^H \mathbf{W} \mathbf{v}_k \right|^2 + \sigma^2} \right)
    \end{equation}
    \begin{equation}
       \text{s.t.} \quad  (q_x, q_y) \in \mathcal{A}
    \end{equation}
    \begin{equation} \label{altitude}
        \quad \quad  z_{\text{min}} \leq q_z \leq z_{\text{max}} 
    \end{equation}
\end{subequations}

where $\mathcal{A}$ is the predefined horizontal coverage region on the (x,y) plane and \eqref{altitude} denotes the altitude constraint. To effectively tackle the problem of maximizing the sum rate, we adopted the gradient ascent-based 3D positioning design. To implement such an approach, it is essential to derive the gradient of the objective function, $R_{\text{sum}}(\mathbf{q})$.

\subsubsection{Gradient Derivation for 3D Location Optimization}
Based on the expression derived above we can write the gradient as  
 \begin{equation}
     \nabla_{\mathbf{q}} R_{\text{sum}} = \sum_{d=1}^D \frac{1}{\ln(2)} \frac{\nabla_{\mathbf{q}} \text{SINR}_d}{1 + \text{SINR}_d}.
 \end{equation}
 
where the gradient of $\text{SINR}_d$ with respect to $\bmq$ can be  obtained using the quotient rule as before
\begin{equation}
    \nabla_{\mathbf{q}} \text{SINR}_d(\bmq) = \frac{\nabla_{\mathbf{q}} N_d(\bmq) \cdot D_d(\bmq) - N_d(\bmq) \cdot \nabla_{\mathbf{q}} D_d(\bmq)}{D_d(\bmq)^2},
\end{equation}
where $ N_d(\bmq)$ and $ D_d(\bmq)$ denote the numerator and the denominator of the $\text{SINR}_d$ at the current position $\bmq$.

The gradient of $ N_d(\bmq) $ can be computed as
\begin{equation}
    \nabla_{\mathbf{q}} N_d(\bmq) = 2 \operatorname{Re} \left( \left( \mathbf{h}_d^H \mathbf{W} \mathbf{v}_d \right)^* \nabla_{\mathbf{q}} \left( \mathbf{h}_d^H \mathbf{W} \mathbf{v}_d \right) \right).
\end{equation}
 
The term \(\nabla_{\mathbf{q}} \mathbf{h}_d^H\) incorporates the gradient of the channel vector. Given the planar array configuration, it is essential to account for the dependence on both the elevation and azimuth angles as well, and not only the path loss, which leads to the following closed-form expression.

\begin{equation}
    \nabla_{\mathbf{q}} \mathbf{h}_d = -\frac{\beta_0}{\|\mathbf{q} - \mathbf{u}_d\|^3} (\mathbf{q} - \mathbf{u}_d) \mathbf{a}(\theta, \phi) + \sqrt{\frac{\beta_0}{\|\mathbf{q} - \mathbf{u}_d\|^2}} \nabla_{\mathbf{q}} \mathbf{a}(\theta, \phi).
\end{equation}

Similarly, for the denominator we can compute the gradient as   
\begin{equation}
     \nabla_{\mathbf{q}} D_d(\bmq)= 2 \operatorname{Re} \left( \left( \mathbf{h}_d(\mathbf{q})^H \mathbf{W} \mathbf{v}_k \right)^* \nabla_{\mathbf{q}} \left( \mathbf{h}_d(\mathbf{q})^H \mathbf{W} \mathbf{v}_k \right) \right).
\end{equation}

\begin{figure*}
\begin{equation} \label{grad_UAV}
\begin{aligned}
    \nabla_{\mathbf{q}} R_{\text{sum}} = & \frac{1}{\ln(2)} \sum_{d=1}^D \frac{1}{1 + \text{SINR}_d} \cdot  \frac{\left( 2 \operatorname{Re} \left( \left( \mathbf{h}_d(\mathbf{q})^H \mathbf{W} \mathbf{v}_d \right)^* \left( -\frac{\beta_0}{\|\mathbf{q} - \mathbf{u}_d\|^3} (\mathbf{q} - \mathbf{u}_d) \mathbf{a}(\theta_d, \phi_d) + \sqrt{\frac{\beta_0}{\|\mathbf{q} - \mathbf{u}_d\|^2}} \mathbf{J}_{\mathbf{a}} \begin{bmatrix} \nabla_{\mathbf{q}} \theta_d \\ \nabla_{\mathbf{q}} \phi_d \end{bmatrix} \right)^H \mathbf{W} \mathbf{v}_d \right)  \right)}{\sum_{k \neq d} \left| \mathbf{h}_d(\mathbf{q})^H \mathbf{W} \mathbf{v}_k \right|^2 + \sigma_d^2} \\ 
    & - \frac{\left| \mathbf{h}_d(\mathbf{q})^H \mathbf{W} \mathbf{v}_d \right|^2 \cdot \sum_{k \neq d} 2 \operatorname{Re} \left( \left( \mathbf{h}_d(\mathbf{q})^H \mathbf{W} \mathbf{v}_k \right)^* \left( -\frac{\beta_0}{\|\mathbf{q} - \mathbf{u}_d\|^3} (\mathbf{q} - \mathbf{u}_d) \mathbf{a}(\theta_d, \phi_d) + \sqrt{\frac{\beta_0}{\|\mathbf{q} - \mathbf{u}_d\|^2}} \mathbf{J}_{\mathbf{a}} \begin{bmatrix} \nabla_{\mathbf{q}} \theta_d \\ \nabla_{\mathbf{q}} \phi_d \end{bmatrix} \right)^H \mathbf{W} \mathbf{v}_k \right)}{ \left(\sum_{k \neq d} \left| \mathbf{h}_d(\mathbf{q})^H \mathbf{W} \mathbf{v}_k \right|^2 + \sigma_d^2 \right)^2}.
\end{aligned}
\end{equation}
\hrulefill
\end{figure*}
Since the gradient of the channel, which depends on path-loss, elevation and azimuth angles, needs a thorough derivation, it is derived in Appendix \ref{appendix_position}. The final expression for the gradient is given in \eqref{grad_UAV}, in which $\mathbf{J}_{\mathbf{a}}$ denotes the Jacobian matrix, $\nabla_{\mathbf{q}} \theta_d$ $\nabla_{\mathbf{q}} \phi_d$ denotes the gradients with respect to the azimuth and elevation angles.

Note that the final gradient incorporates.
 
\begin{itemize}
    \item The term $ -\frac{\beta_0}{\|\mathbf{q} - \mathbf{u}_d\|^3} (\mathbf{q} - \mathbf{u}_d) \mathbf{a}(\theta_d, \phi_d) $ accounting for the gradient of the path loss.
    \item The term $ \mathbf{J}_{\mathbf{a}} \begin{bmatrix} \nabla_{\mathbf{q}} \theta_d \\ \nabla_{\mathbf{q}} \phi_d \end{bmatrix} $ captures the variation of the UPA steering vector with respect to the UAV's position, thus takes into account also how the signal is combined at the UAV.
\end{itemize}
Given the gradient, the following iteration rule can be applied to update the trajectory 
\begin{equation}
    \mathbf{q}(n+1) = \mathbf{q}(t) + \mu_q \nabla_{\mathbf{q}} R_{\text{sum}},
\end{equation}
where $\mu_q$ denotes the learning step size.
To ensure that the UAV's trajectory remains within the feasible operational space, we project it in the constrained 3D region. The altitude constraint ensures that the UAV's altitude $ q_z(n+1) $ remains within the predefined vertical bounds $ z_{\text{min}} $ and $ z_{\text{max}} $. Mathematically, this is achieved through the following projection
\begin{equation}
q_z(n+1) \leftarrow \max\left(z_{\text{min}}, \min(z_{\text{max}}, q_z(n+1))\right).
\end{equation}

For the horizontal coverage constraint, the updated horizontal coordinates $ (q_x(n+1), q_y(n+1)) $ are projected onto the operational area $ \mathcal{A} $. If $ \mathcal{A} $ is a rectangular region defined by bounds $ [x_{\text{min}}, x_{\text{max}}] $ and $ [y_{\text{min}}, y_{\text{max}}] $, the projection is
\begin{equation}
q_x(n+1) \leftarrow \max\left(x_{\text{min}}, \min(x_{\text{max}}, q_x(n+1))\right),
\end{equation}
\begin{equation}
q_y(n+1) \leftarrow \max\left(y_{\text{min}}, \min(y_{\text{max}}, q_y(n+1))\right).
\end{equation}

If $ \mathcal{A} $ is a circular region centered at $ \mathbf{c} = [x_c, y_c]^T $ with radius $ r $, the projection is
\begin{equation}
\mathbf{q}_{\text{proj}} = 
\begin{cases}
\mathbf{q}(n+1), & \text{if } \|\mathbf{q}_{\text{xy}}(n+1) - \mathbf{c}\| \leq r, \\[6pt]
\mathbf{c} + r \frac{\mathbf{q}_{\text{xy}}(n+1) - \mathbf{c}}{\|\mathbf{q}_{\text{xy}}(n+1) - \mathbf{c}\|}, & \text{otherwise}.
\end{cases}
\end{equation}
By iteratively applying these constraints during the UAV's 3D position optimization, the UAV's position updates remain both feasible and compliant with physical and operational limits, ultimately converging to a 3D position that maximizes the system's performance.
In practice, it is important to note that the UAV may also be subject to velocity constraints while transitioning between points. To account for such constraints, the gradient can be normalized, and the step size $\mu_q$ can be appropriately selected to ensure that the maximum displacement of the UAV at each step remains feasible.

\begin{algorithm}[!t]
\caption{Joint Optimization of UAV 3D Position, Holographic Beamforming, and Digital Beamforming}
\begin{enumerate}
    \item \textbf{Initialization}: 
    \begin{itemize}
        \item Initialize the UAV position, holographic beamformer $\bmW$, and the digital beamformer $\bmV$.
        \item Specify the learning step sizes $\mu_q$ and $\eta$.
        \item Initialize iteration counter $t = 0$ and sum rate $R_{\text{prev}} = 0$.
    \end{itemize}
    
    \item \textbf{While}
    \begin{enumerate}
        \item  {Optimize} $\mathbf{W}$ with Algorithm \ref{holographic_alg}.
        \item  {Optimize} $\mathbf{V}$ with ZF.
        
        \item {Update UAV Position as}  
        \[
        \mathbf{q}(n+1) = \mathbf{q}(t) + \mu_q \nabla_{\mathbf{q}} R_{\text{sum}}.
        \]
        
        \item {Projection Step for UAV Position:}
        \begin{itemize}
            \item Enforce altitude constraints:
            \[
            q_z(n+1) \leftarrow \max\left(z_{\text{min}}, \min(z_{\text{max}}, q_z(n+1))\right).
            \]
            \item Enforce horizontal coverage constraints:
            \[
            (q_x(n+1), q_y(n+1)) \in \mathcal{A}
             \]
        \end{itemize}
        
        \item \textbf{Convergence Check:} Compute the difference in sum rate:
        \[
        \Delta R = \left| R_{\text{sum}}^{(n+1)} - R_{\text{sum}}^{(t)} \right|.
        \]
        If $\Delta R < \epsilon_{\text{tol}}$, terminate; otherwise, update $R_{\text{prev}} \leftarrow R_{\text{sum}}^{(n+1)}$ and increment $t \leftarrow t + 1$.
    \end{enumerate}
    
    \item \textbf{Output:} $\bmW$, $\bmV$ and $\bmq$.
\end{enumerate}
\end{algorithm}

\subsection{Convergence Analysis}

The proposed algorithm ensures convergence for the joint optimization of the UAV’s 3D position, holographic beamforming, and digital beamforming by leveraging alternating optimization and the monotonic improvement of the sum rate. The problem is divided into three subproblems, with one optimized per iteration while the others are fixed.

At each iteration \(t\), the system sum rate \(R_{\text{sum}}(t)\) satisfies:
\begin{equation}
    R_{\text{sum}}(t+1) \geq R_{\text{sum}}(t),
\end{equation}
ensuring non-decreasing behavior. As \(R_{\text{sum}}(t)\) is bounded above, the sequence converges to an asymptotic value \(R_{\text{sum}}^*\):
\begin{equation}
    \lim_{t \to \infty} R_{\text{sum}}(t) = R_{\text{sum}}^*.
\end{equation}

Holographic beamformer \(\mathbf{W}\) is optimized using gradient ascent. After each update, the weights are projected onto the feasible set \([0,1]^M\), ensuring physical constraints. The sum rate \(R_{\text{sum}}(\mathbf{w})\) improves monotonically at every iteration, guaranteeing convergence to a local optimum.

Digital beamforming is optimized using zero-forcing (ZF), which directly computes the beamforming matrix to eliminate inter-user interference. This step provides an exact solution and converges immediately without approximation errors.
The UAV position \(\mathbf{q}\) is updated iteratively using gradient ascent:
\[
\mathbf{q}(t+1) = \mathbf{q}(t) + \mu_q \nabla_{\mathbf{q}} R_{\text{sum}},
\]
where \(\mu_q\) is the step size. The UAV position is projected onto the operational space defined by altitude constraints \([z_{\text{min}}, z_{\text{max}}]\) and horizontal coverage constraints \(\mathcal{A}\), ensuring feasibility. This step guarantees monotonic improvement in \(R_{\text{sum}}(\mathbf{q})\) and convergence to a local optimum.

\begin{figure} [t]
    \centering
    \includegraphics[width=8cm,height=5.5cm]{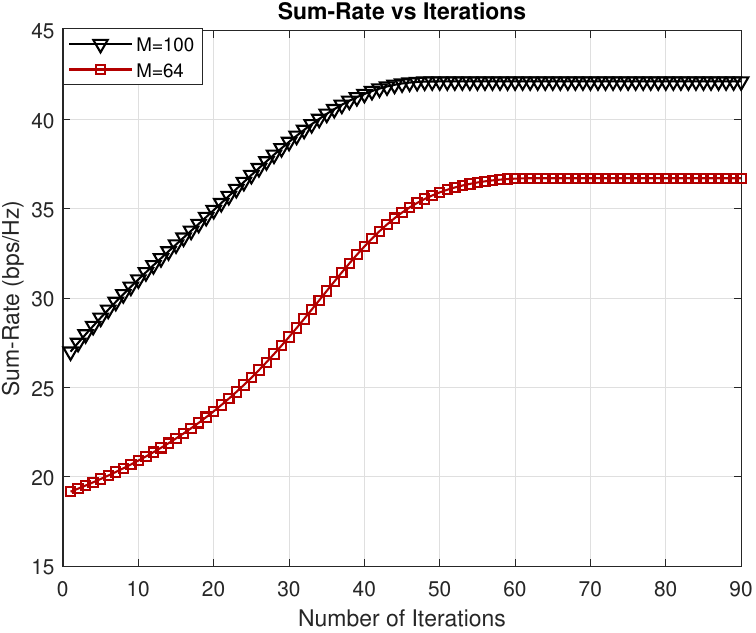}
    \caption{Convergence of the proposed algorithm with $M=64$ and $M=100$ at SNR=$30$dB.}
    \label{convergenza}
\end{figure}

The algorithm iteratively solves these subproblems, ensuring monotonic improvement of the sum rate and convergence to a locally optimal solution. However, due to the nonconvex nature of the problem, the solution depends on the initialization of the UAV position, holographic beamforming weights, and digital beamforming vectors. A typical convergence behaviour of the proposed algorithm with different holographic beamformer sizes is presented in Figure \ref{convergenza}.

\subsection{Complexity Analysis}

In this section, we analyze the complexity of the proposed joint optimization algorithm.

The optimization step for holographic beamforming involves updating the holographic weights to maximize the sum rate of the system. For each holographic element $ w_m $, the gradient $ \frac{\partial R_{\text{sum}}}{\partial w_m} $ must be computed. The computation involves evaluating the signal-to-interference-plus-noise ratio (SINR) for $ D $ users, leading to a complexity of $\mathcal{O}(M \cdot K \cdot D) $
per element. The total complexity of gradient computation across all $ M $ elements is therefore $ \mathcal{O}(M^2 \cdot K \cdot D) $. Updating and projecting the $ M $ weights back into the feasible range contributes an additional $ \mathcal{O}(M) $. Thus, the total complexity of holographic beamforming per iteration of joint algorithm is  $\mathcal{O}( T_w M^2 \cdot K \cdot D)$, where $T_w$ denotes the iterations required by Algorithm $1$ to optimally update the holographic weights until convergence.
The digital beamforming optimization step involves computing the ZF beamforming matrix. The effective channel matrix $ \mathbf{H}^H(\mathbf{q}) \mathbf{W} $ of size $ D \times K $ is pseudo-inverted to obtain the ZF solution. Furthermore, the normalization of the power of the beamforming vectors requires $ \mathcal{O}(D) $. The total complexity of the digital beamforming optimization step is therefore $\mathcal{O}(K^3 + K^2 \cdot D)$. The optimization step for UAV 3D positioning involves computing the gradient of the sum rate of the system, $ \nabla_{\mathbf{q}} R_{\text{sum}} $, with respect to UAV position $ \mathbf{q} $. This gradient computation requires evaluating the SINR gradient for $ D $ users. For each user, it results to be $ \mathcal{O}(M \cdot K + M^2) $, and the total complexity is given as $\mathcal{O}(D \cdot M \cdot K + D \cdot M^2).$
Updating the UAV position and projecting it onto the feasible region is computationally inexpensive, adding negligible complexity.  Combining the complexities of the three steps, the total complexity of the algorithm per iteration is:
\begin{equation}
\mathcal{O}( T_w M^2 \cdot K \cdot D + K^3 + K^2 \cdot D + D \cdot M^2 + D \cdot M \cdot K).
\end{equation}

For large $ M $, the term $ T_w\cdot M^2 \cdot K \cdot D $ dominates the complexity due to the quadratic dependence on the number of holographic elements.

\begin{figure*}
    \centering
 \begin{minipage}{0.49\textwidth}
      \centering
    \includegraphics[width=\linewidth]{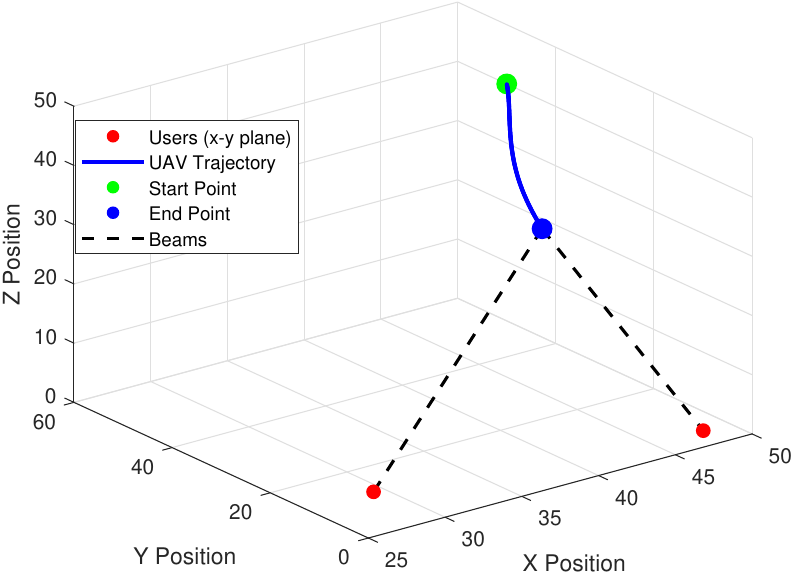}
    \caption{3D Position optimization of the UAV for holographic communications with $D =2$ users.}
    \label{fig3}
\end{minipage}  
      \begin{minipage}{0.49\textwidth}
      \centering
    \includegraphics[width=\linewidth]{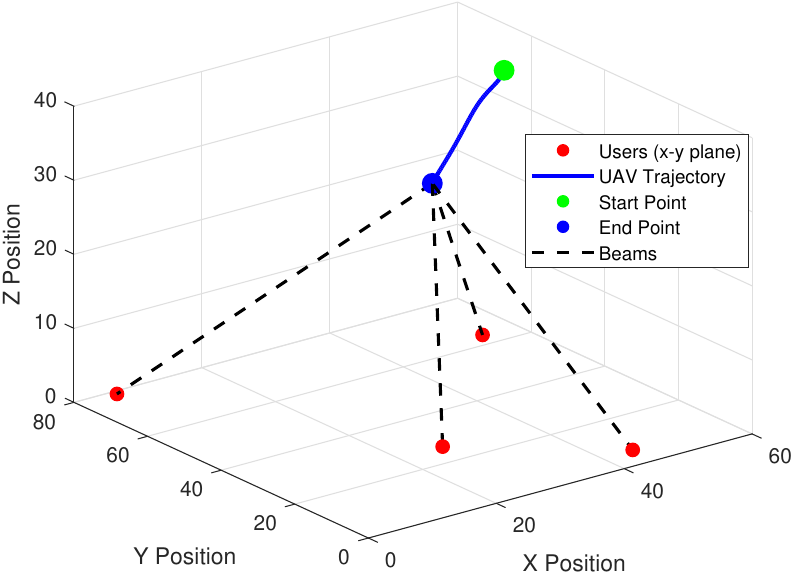}
    \caption{3D Position optimization of the UAV for holographic communications with $D =4$ users.}
    \label{fig4}
    \end{minipage}  
\end{figure*}
\section{Simulation Results} \label{section_4}
In this section, we present simulation results to evaluate the performance of the proposed algorithm for joint holographic beamforming and 3D UAV positioning. The simulations are conducted over 100 realizations to capture the variability in the system's performance under different conditions. The system consists of $2$ or $4$ users which are randomly placed on the surface limited by the following  $[0, 100\text{m}]$ along the x-axis and $ [0, 100\text{m}]$ along the y-axis. The UAV is assumed to have $6$ RF chains, and its initial position is at the coordinates $[50, 50, 40]$. The UAV's position is updated iteratively with constraints within a defined 3D space. The holographic beamformer employs iterative amplitude optimization, for which the convergence threshold of $ \epsilon = 10^{-5} $ and the learning rate of $ \eta = 0.01 $ are chosen. For UAV movement, the gradient is normalized and the learning rate is set as $\mu_q=2$, which ensures that the UAV moves $2$ m at each iteration to find the optimal 3D position. The carrier frequency of $30$ GHz is assumed. Recall that the propagation vector in the free space is represented by $ \mathbf{k}_f $, while the propagation vector in the RHS is denoted as $ \mathbf{k}_s $. According to electromagnetism, the magnitude of $ \mathbf{k}_s $ is related to $ \mathbf{k}_f $ by the equation $|\mathbf{k}_s| = \sqrt{\epsilon_r} |\mathbf{k}_f|$, where $ \epsilon_r $ is the relative permittivity of the RHS substrate, typically around $3$. These parameters are set with typical values used in the literature as $|\mathbf{k}_s| = 200 \pi $ and $|\mathbf{k}_f| = 200 \sqrt{3} \pi$ \cite{deng2022reconfigurable}, and the spacing between the elements is set as $\lambda/3$. 

We define the SNR of the system as the transmit SNR, that is, $ \text{SNR} = P_{\text{max}}/\sigma_b^2,$ with $\sigma_b^2$ denoting the noise variance at the base station. The base station and the users are assumed to have the same level of noise variance, i.e., $\sigma_b = \sigma_d, \forall d$, which is set to meet the desirable SNR values given the total transmit power. The total transmit power at the base station is set to $ P_{\text{max}} = 1$W, and the noise power is chosen to meet the desirable SNR. The UAV's position is constrained to a 3D space with minimum and maximum values defined as $q_{\text{min}} = [0, 0, 10]$ for the minimum position, and $q_{\text{max}} = [100, 100, 50]$ for the maximum position, where $x$, $y$, and $z$ represent the coordinates of the UAV. The RHS is assumed to be either of size $10 \times 10= 100$ or $8 \times 8= 64$ for holographic weight-controlled beamforming, unless otherwise stated. Finally, the channel gain $\beta_0$ is set to $1$.

To evaluate the performance of the proposed method, a benchmark scheme is defined, in which digital beamformers and 3D UAV position are optimized, while holographic beamformer amplitudes are randomly selected within the interval $[0, 1]$. We will refer to such a scheme as a \emph{Benchmark} in the following, and refer to the joint optimization developed in this work as \emph{Proposed}.

Figure \ref{fig3} illustrates the simulated trajectory of a UAV in the case of two users to find the optimal position as it moves from the starting point to the endpoint under the proposed algorithm designed to maximize the system's sum rate. The UAV begins at the starting point $[50, 50, 40]$, marked by the green point, and converges to the endpoint, represented by the blue point, corresponding to the optimal position that provides the maximum sum rate. The trajectory, shown as a curved blue line, reflects the UAV's dynamic adjustments to optimize coverage and connectivity by balancing altitude and horizontal movement. Red points on the x-y plane indicate the user locations, while dashed black lines represent communication beams between the UAV and users. Figure \ref{fig4} shows the UAV trajectory to find the optimal 3D location in the case of four users, starting from the same initial 3D position. We can see the UAV's trajectory adapts to accommodate the increased complexity of the network while still maximizing the system's sum rate. Similar to the two users case, the UAV, starting from the initial point, descends and maneuvers through the space, ultimately converging at the blue point, representing the optimal position. 


From the figures \ref{fig3} and \ref{fig4}, we can observe that in the two-user scenario, the optimal position of the UAV is determined so that the distances from the UAV to both users are approximately equal. This ensures that the path loss is balanced, providing a more or less equal sum-rate contribution from each user. In contrast, the four-user case presents a more complex scenario, as the UAV needs to accommodate the communication requirements of multiple users with diverse spatial distributions. Here, the optimal position shifts towards the region with a higher density of users to maximize the system's sum rate. 
 
\begin{figure*}
    \centering
 \begin{minipage}{0.49\textwidth}
     \centering
   \includegraphics[width=0.9\linewidth]{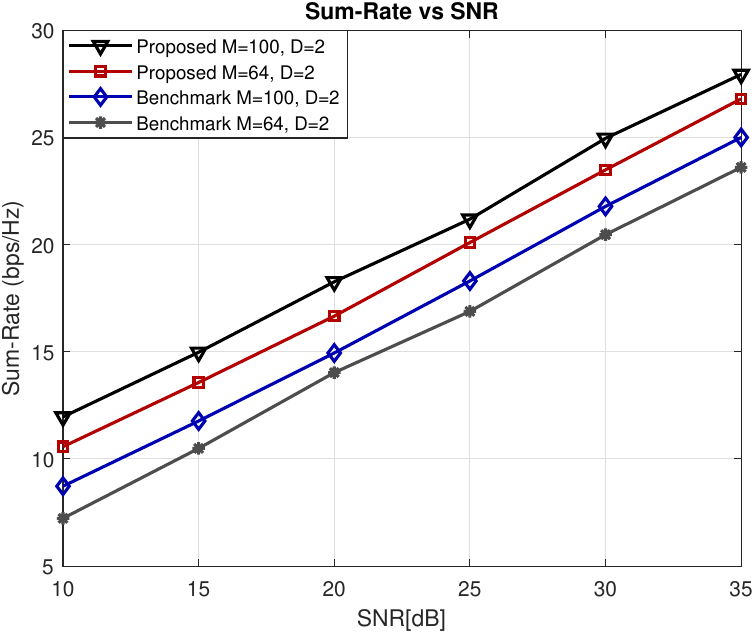}
   \caption{Sum-rate of holographic UAV communications as a function of SNR with $D = 2$.}
   \label{fig5}
\end{minipage}  
      \begin{minipage}{0.49\textwidth}
       \centering
   \includegraphics[width=0.9\linewidth]{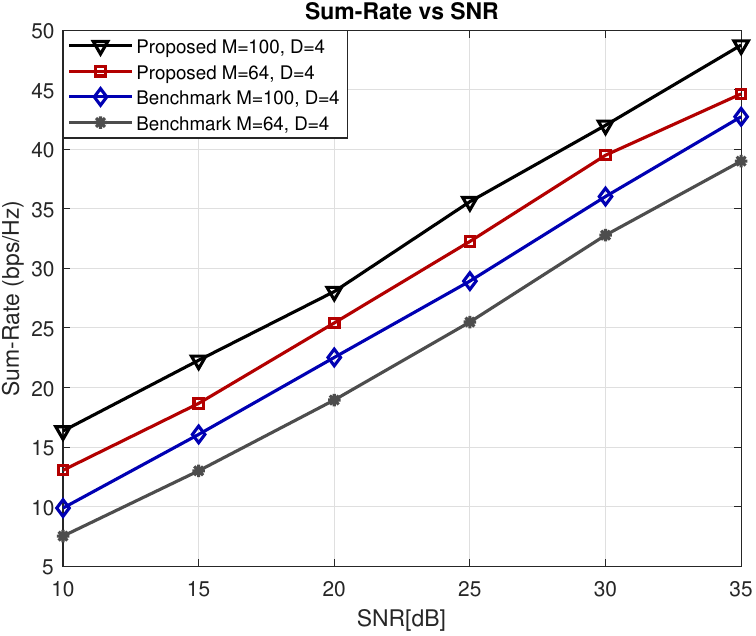}
   \caption{Sum-rate of holographic UAV communications as a function of SNR with $D = 4$.}
    \label{fig6}
    \end{minipage}  
\end{figure*}

Figure \ref{fig5} demonstrates the performance of the proposed algorithm for joint holographic beamforming and 3D UAV positioning compared to the benchmark scheme in a system with two users (\(D = 2\)) and holographic surface sizes \(M = 100\) and \(M = 64\). The results are plotted as a function of the SNR. We can see that the proposed method consistently outperforms the benchmark scheme across all SNR levels, with the performance gap becoming more pronounced at higher SNRs. The results also show that larger holographic surface sizes (\(M = 100\)) yield higher sum rates for both methods, as the larger surface provides enhanced holographic beamforming resolution and improved signal alignment, resulting in more efficient holographic UAV communication. The benchmark scheme, which relies on randomly selected holographic beamformer amplitudes, achieves significantly lower sum rates at all SNRs, emphasizing the inefficiency of random amplitude selection to optimally exploit the potential of holographic beamforming. Figure \ref{fig4} shows the case of four users (\(D = 4\)) and provides insight into how the performance of the system evolves when the number of users increases compared to the case of two users. The proposed algorithm again outperforms the benchmark scheme in all SNR levels for both holographic surface sizes (\(M = 100\) and \(M = 64\)).  A noticeable observation in the four-user case is that the sum-rate values for all methods have increased compared to the two-user scenario. This is due to the additional degrees of freedom in optimizing the UAV's position and holographic beamforming to serve a larger number of users. We can see that the proposed algorithm effectively adapts to the increased demand by leveraging the joint optimization approach, ensuring that the system remains scalable and efficient in managing the interference, even when the number of users increases.

\begin{figure*}
    \centering
 \begin{minipage}{0.49\textwidth}
      \centering
    \includegraphics[width=0.9\linewidth]{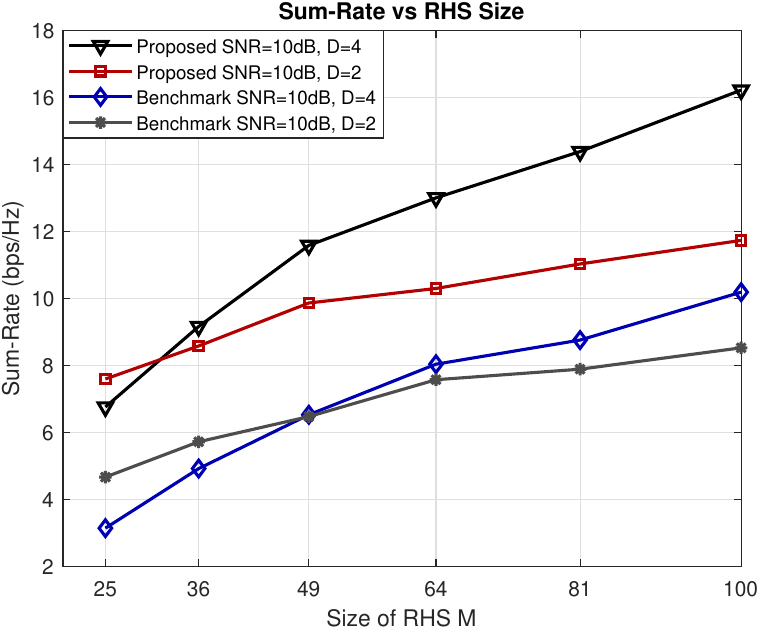}
    \caption{Sum-rate of holographic UAV communications in comparison with the size of the RHS at SNR$=10$dB.}
    \label{fig7}
\end{minipage}  
      \begin{minipage}{0.49\textwidth}
      \centering
    \includegraphics[width=0.9\linewidth]{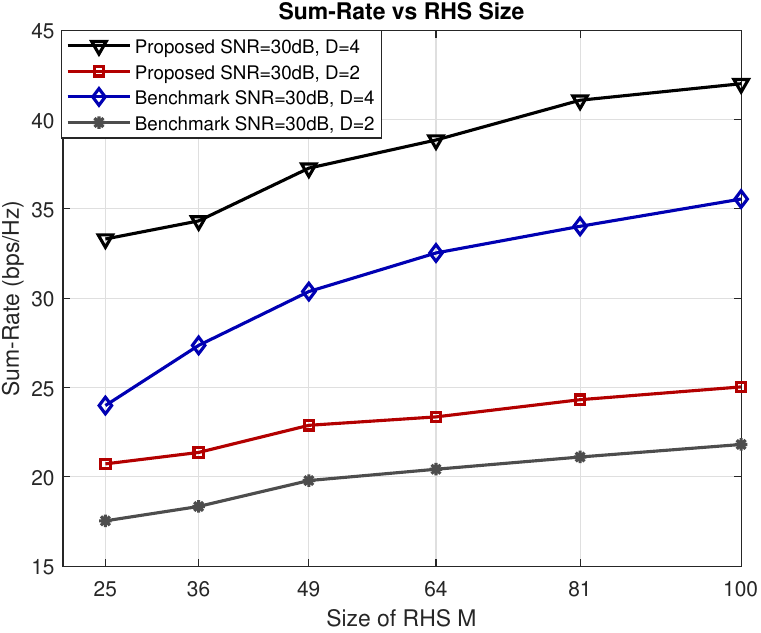}
    \caption{Sum-rate of holographic UAV communications in comparison with the size of the RHS at SNR$=30$dB.}
    \label{fig8}
    \end{minipage}  
\end{figure*}

Figure \ref{fig7} compares the performance of the proposed algorithm and the benchmark scheme as a function of the holographic beamformer size (\(M\)) at an SNR of 10 dB for systems with two (\(D = 2\)) and four users (\(D = 4\)). The results highlight several important observations about system behavior under low SNR conditions and varying RHS sizes. Firstly, we can see that at a smaller RHS size, the system struggles to manage interference effectively, resulting in limited sum-rate performance, especially in the four-user case. This is attributed to the inherent challenges of zero-forcing beamforming at low SNR, which is known to perform suboptimally under such conditions. The combination of small RHS size and low SNR exacerbates these issues, as the reduced hybrid holographic beamforming resolution limits the system's ability to spatially separate user signals, leading to increased interference and degraded performance. We can see that when the RHS size is very small, the system achieves a better sum rate in the case of two users compared to the four users case. As the RHS size increases, the performance of both the proposed method and the benchmark improves due to the enhanced holographic beamforming capabilities provided by larger RHS sizes, while the performance of the ZF digital beamforming remains limited. The proposed algorithm demonstrates a significant advantage over the benchmark across all RHS sizes and the number of user configurations. This is particularly emphasized for the four users, reflected in the steeper growth of the sum rate with \(M\) for the proposed method compared to the benchmark, showcasing its robustness in managing interference and adapting to the demands of the system. In the two-user case, the impact of interference is less pronounced, and the system performs well compared to the four-user case at a very small RHS size. However, even in this scenario, the proposed method consistently outperforms the benchmark at each RHS size, highlighting the effectiveness of the optimization approach. These results underscore the importance of large RHS sizes in improving the system's performance even at extremely low SNR. At the same time, the potential of digital beamforming remains limited due to a limited number of RF chains and low SNR.

Finally, figure \ref{fig8} shows the impact of increasing the holographic beamformer size (\(M\)) on the sum-rate for systems with two (\(D = 2\)) and four users (\(D = 4\))  at an SNR of $30$ dB. We can see that at high SNR, the sum-rate for both the proposed algorithm and the benchmark scheme improves as \(M\) increases.  The proposed algorithm demonstrates superior performance compared to the benchmark scheme across all \(M\) values and user configurations. The advantage of the proposed algorithm is particularly pronounced in the four-user case (\(D = 4\)), where the system faces greater complexity in handling interference and optimizing the UAV’s position. In contrast to the low-SNR scenario, it is clearly visible that the SNR mitigates the limitations of zero-forcing beamforming. This results in improved interference suppression, which allows the system to achieve better performance.  For the two-user case (\(D = 2\)), the performance trends are similar, with the proposed algorithm consistently achieving higher sum rates than the benchmark. However, the relative improvement is less dramatic than in the four-user case, as the reduced number of users leads to reduced spatial multiplexing of the data streams.

The simulation results demonstrated the effectiveness of the proposed algorithm for joint holographic beamforming and 3D UAV positioning in optimizing the system's sum rate under various configurations and conditions. The proposed algorithm consistently outperforms the benchmark scheme in different SNR levels, holographic beamformer sizes, and user configurations, highlighting its robustness and adaptability. 

Analysis of UAV trajectories shows that the proposed algorithm enables dynamic adjustments to optimize coverage and connectivity. In the two-user scenario, the UAV converges to an optimal position that balances path loss for both users, ensuring equitable contributions to the sum rate. In contrast, the four-user case presents a more complex spatial distribution, requiring the UAV to prioritize regions with higher user density to maximize the system's performance. These results show the scalability of the proposed method in handling increasing user demands.

The impact of SNR is evident in the simulation results. At low SNR, the system struggles with interference management, particularly when the RHS size is small. This is further exacerbated by the inherent limitations of zero-force beamforming, which performs suboptimally at low SNR. However, the proposed algorithm demonstrates resilience, achieving significantly higher sum rates compared to the benchmark scheme. At high SNR, the system benefits from enhanced interference suppression, allowing both methods to perform better. However, the proposed algorithm significantly outperforms the benchmark, particularly in the case of four users, where the system faces greater complexity.

The results also emphasize the critical role of RHS size. Larger holographic beamformer sizes improve beamforming resolution and signal alignment, resulting in higher sum rates for both two-user and four-user scenarios. This improvement is particularly pronounced in the four-user case, where managing interference becomes increasingly challenging as the number of users increases. The proposed algorithm leverages these enhancements more effectively than the benchmark, further widening the performance gap.

In conclusion, the proposed algorithm demonstrates strong performance in all scenarios tested, leveraging iterative optimization of holographic beamformer amplitudes and UAV positioning. These findings validate the method's ability to maximize the system sum rate, adapt to varying user distributions, and manage interference effectively, even in challenging multiuser and low-SNR environments.

\section{Conclusions} \label{section_5}
 
This paper introduced a novel framework for multi-user holographic UAV communications, leveraging the unique capabilities of RHSs for joint holographic beamforming and 3D UAV positioning. The study addressed the complex optimization problem of maximizing the system's sum rate by iteratively adjusting the UAV's 3D position as well as the analog and digital beamformers. A hybrid approach combining ZF digital beamforming, iterative holographic amplitude optimization, and gradient-based 3D UAV positioning was proposed, ensuring convergence and practical feasibility, while achieving significant system performance gains. The proposed results aim to promote a new era of UAV communications, consisting of a lightweight, low-cost and energy-efficient design, making it particularly well-suited for UAVs, where constraints on power, payload, and operational range are critical. The effectiveness of the proposed approach was validated through comprehensive numerical evaluations, which demonstrated its superior performance compared to the benchmark scheme across various systems configurations.

\begin{appendices}
\section{Gradient Derivation for Holographic Beamforming} \label{appendix_HBF}

The gradient derivation for holographic beamforming involves calculating the partial derivatives of the numerator $ N_d(w_m) $ and the denominator $ D_d(w_m) $ of the SINR with respect to the weight $ w_m $ of the RHS elements. 

The numerator $ N_d(w_m) $ is expressed as:
\begin{equation}
    N_d(w_m) = \left|u_d(\mathbf{w})\right|^2,
\end{equation}
where:
\begin{equation}
    u_d(\mathbf{w}) = \sum_{m=1}^M w_m h_{d,m}^*(\mathbf{q}) (\mathbf{\Phi} \mathbf{v}_d)_m.
\end{equation}

To compute the derivative of $ N_d(w_m) $, we apply the product rule:
\begin{equation}
    \frac{\partial N_d}{\partial w_m} = \left(\frac{\partial u_d}{\partial w_m}\right) u_d^* + u_d \left(\frac{\partial u_d^*}{\partial w_m}\right).
\end{equation}

The derivative of $ u_d(\mathbf{w}) $ with respect to $ w_m $ is:
\begin{equation}
    \frac{\partial u_d}{\partial w_m} = h_{d,m}^*(\mathbf{q}) (\mathbf{\Phi} \mathbf{v}_d)_m.
\end{equation}

Substituting this back into the expression for $ \frac{\partial N_d}{\partial w_m} $, we have:
\begin{equation}
    \frac{\partial N_d}{\partial w_m} = h_{d,m}^*(\mathbf{q}) (\mathbf{\Phi} \mathbf{v}_d)_m u_d^* + h_{d,m}(\mathbf{q}) (\mathbf{\Phi} \mathbf{v}_d)_m^* u_d.
\end{equation}

For the denominator $ D_d(w_m) $, it is expanded as:
\begin{equation}
    D_d(w_m) = \sum_{k \neq d} \left|u_{d,k}(\mathbf{w})\right|^2 + \sigma^2,
\end{equation}
where:
\begin{equation}
    u_{d,k}(\mathbf{w}) = \sum_{m=1}^M w_m h_{d,m}^*(\mathbf{q}) (\mathbf{\Phi} \mathbf{v}_k)_m.
\end{equation}

To differentiate $ |u_{d,k}(\mathbf{w})|^2 $, we use:
\begin{equation}
    \frac{\partial |u_{d,k}(\mathbf{w})|^2}{\partial w_m} = u_{d,k}^* h_{d,m}^*(\mathbf{q}) (\mathbf{\Phi} \mathbf{v}_k)_m + u_{d,k} h_{d,m}(\mathbf{q}) (\mathbf{\Phi} \mathbf{v}_k)_m^*.
\end{equation}

By summing over all interfering users $ k \neq d $, the derivative of $ D_d(w_m) $ becomes:
\begin{equation}
    \frac{\partial D_d}{\partial w_m} = \sum_{k \neq d} \bigl[u_{d,k}^* h_{d,m}^*(\mathbf{q}) (\mathbf{\Phi} \mathbf{v}_k)_m + u_{d,k} h_{d,m}(\mathbf{q}) (\mathbf{\Phi} \mathbf{v}_k)_m^*\bigr].
\end{equation}

Finally, substituting the expressions for $ \frac{\partial N_d}{\partial w_m} $ and $ \frac{\partial D_d}{\partial w_m} $ into $ \frac{\partial \text{SINR}_d}{\partial w_m} $, we obtain:
\begin{align}
    \frac{\partial \text{SINR}_d}{\partial w_m} = \frac{\bigl[h_{d,m}^*(\mathbf{q})(\mathbf{\Phi} \mathbf{v}_d)_m u_d^* + h_{d,m}(\mathbf{q})(\mathbf{\Phi} \mathbf{v}_d)_m^* u_d\bigr] D_d}{D_d^2} \nonumber \\
    - \frac{N_d \sum_{k \neq d} \bigl[u_{d,k}^* h_{d,m}^*(\mathbf{q})(\mathbf{\Phi} \mathbf{v}_k)_m + u_{d,k} h_{d,m}(\mathbf{q})(\mathbf{\Phi} \mathbf{v}_k)_m^*\bigr]}{D_d^2}.
\end{align}
 
Finally, substituting this result into the gradient of the sum rate, we get the final expression for the gradient of the sum-rate with respect to the weight $w_m$.

\section{Gradient Derivation for 3D Position Optimization} \label{appendix_position}

This appendix provides the derivation of the gradient of the term $ \mathbf{h}_d^H \mathbf{W} \mathbf{v}_d $ with respect to the UAV's position $ \mathbf{q} $. Here, $ \mathbf{h}_d(\mathbf{q}) $ represents the channel vector between the UAV and user $ d $, $ \mathbf{W} $ denotes the holographic beamforming matrix, and $ \mathbf{v}_d $ is the digital beamforming vector. The gradient is expressed as:
\begin{equation}
\nabla_{\mathbf{q}} \left( \mathbf{h}_d^H \mathbf{W} \mathbf{v}_d \right) = \left( \nabla_{\mathbf{q}} \mathbf{h}_d \right)^H \mathbf{W} \mathbf{v}_d.
\end{equation}

The channel vector is modelled as:
\begin{equation}
\mathbf{h}_d(\mathbf{q}) = \sqrt{\frac{\beta_0}{\|\mathbf{q} - \mathbf{u}_d\|^2}} \mathbf{a}(\theta_d, \phi_d),
\end{equation}
where $ \beta_0 $ is the reference channel gain, $ \|\mathbf{q} - \mathbf{u}_d\| $ is the distance between the UAV and user $ d $, and $ \mathbf{a}(\theta_d, \phi_d) $ is the uniform planar array (UPA) steering vector.

To compute $ \nabla_{\mathbf{q}} \mathbf{h}_d $, we decompose it into the gradient of the path loss term and the gradient of the steering vector. The path loss term, $ \sqrt{\frac{\beta_0}{\|\mathbf{q} - \mathbf{u}_d\|^2}} $, has a gradient given by:
\begin{equation}
\nabla_{\mathbf{q}} \left( \sqrt{\frac{\beta_0}{\|\mathbf{q} - \mathbf{u}_d\|^2}} \right) = -\frac{\beta_0}{\|\mathbf{q} - \mathbf{u}_d\|^3} (\mathbf{q} - \mathbf{u}_d).
\end{equation}

For the steering vector, $ \mathbf{a}(\theta_d, \phi_d) $, which depends on the azimuth and elevation angles $ \phi_d $ and $ \theta_d $, we express the gradient as:
\begin{equation}
\nabla_{\mathbf{q}} \mathbf{a}(\theta_d, \phi_d) = \mathbf{J}_{\mathbf{a}} \begin{bmatrix} \nabla_{\mathbf{q}} \theta_d \\ \nabla_{\mathbf{q}} \phi_d \end{bmatrix},
\end{equation}
where $ \mathbf{J}_{\mathbf{a}} $ is the Jacobian matrix of $ \mathbf{a} $ with respect to $ \theta_d $ and $ \phi_d $, which is separately derived in Appendix \ref{appendix_jacob}. The angles are functions of $ \mathbf{q} $, and their gradients are given by $ \nabla_{\mathbf{q}} \theta_d $ and $ \nabla_{\mathbf{q}} \phi_d $, with explicit expressions derived geometrically:
\begin{equation}
\nabla_{\mathbf{q}} \theta_d = \frac{1}{\|\mathbf{q} - \mathbf{u}_d\|} 
\begin{bmatrix}
-\cos\theta_d \cos\phi_d \\
-\cos\theta_d \sin\phi_d \\
\sin\theta_d
\end{bmatrix},
\end{equation}
\begin{equation}
\nabla_{\mathbf{q}} \phi_d = \frac{1}{\|\mathbf{q} - \mathbf{u}_d\|^2} 
\begin{bmatrix}
-\sin\phi_d \\
\cos\phi_d \\
0
\end{bmatrix}.
\end{equation}

Combining these components, the gradient of the channel vector becomes:
\begin{equation}
\nabla_{\mathbf{q}} \mathbf{h}_d = -\frac{\beta_0}{\|\mathbf{q} - \mathbf{u}_d\|^3} (\mathbf{q} - \mathbf{u}_d) \mathbf{a}(\theta_d, \phi_d)^H + \sqrt{\frac{\beta_0}{\|\mathbf{q} - \mathbf{u}_d\|^2}} \mathbf{J}_{\mathbf{a}} \begin{bmatrix} \nabla_{\mathbf{q}} \theta_d \\ \nabla_{\mathbf{q}} \phi_d \end{bmatrix}^H.
\end{equation}

Substituting this result, the gradient of $ \mathbf{h}_d^H \mathbf{W} \mathbf{v}_d $ is:
\begin{equation}
\begin{aligned}
      \nabla_{\mathbf{q}} \left( \mathbf{h}_d^H \mathbf{W} \mathbf{v}_d \right) = & \Bigg[ 
      -\frac{\beta_0}{\|\mathbf{q} - \mathbf{u}_d\|^3} (\mathbf{q} - \mathbf{u}_d) \mathbf{a}(\theta_d, \phi_d)^H \\
      & \quad + \sqrt{\frac{\beta_0}{\|\mathbf{q} - \mathbf{u}_d\|^2}} \mathbf{J}_{\mathbf{a}} 
      \begin{bmatrix} 
          \nabla_{\mathbf{q}} \theta_d \\ 
          \nabla_{\mathbf{q}} \phi_d 
      \end{bmatrix}^H
      \Bigg] \mathbf{W} \mathbf{v}_d.
\end{aligned}
\end{equation}

This formulation provides a comprehensive expression for the gradient, accounting for both the path loss and steering vector variations due to changes in the UAV’s position $ \mathbf{q} $.

The gradient of the denominator can be derived similarly which leads to the following expression 
\begin{equation}
   \begin{aligned}
       \nabla_{\mathbf{q}} D_d(\bmq) = & \sum_{k \neq d} 2 \operatorname{Re} \Bigg( \left( \mathbf{h}_d^H \mathbf{W} \mathbf{v}_k \right)^* \Bigg[
       -\frac{\beta_0}{\|\mathbf{q} - \mathbf{u}_d\|^3} (\mathbf{q} - \mathbf{u}_d) \\ & \mathbf{a}(\theta_d, \phi_d)^H 
       + \sqrt{\frac{\beta_0}{\|\mathbf{q} - \mathbf{u}_d\|^2}} \mathbf{J}_{\mathbf{a}} 
       \begin{bmatrix} 
           \nabla_{\mathbf{q}} \theta_d \\ 
           \nabla_{\mathbf{q}} \phi_d 
       \end{bmatrix}^H
       \Bigg] \mathbf{W} \mathbf{v}_k \Bigg).
   \end{aligned}
\end{equation}

\section{Derivation of the Jacobian Matrix} \label{appendix_jacob}

The Jacobian matrix \(\mathbf{J}_{\mathbf{a}}\) of the steering vector \(\mathbf{a}(\theta, \phi)\) is derived to capture the sensitivity of the steering vector to variations in the elevation angle \(\theta\) and the azimuth angle \(\phi\). The steering vector for a uniform planar array (UPA) is expressed as:
\begin{equation}
\mathbf{a}(\theta, \phi) = \mathbf{a}_y(\theta, \phi) \otimes \mathbf{a}_x(\theta, \phi),
\end{equation}
where \(\mathbf{a}_x(\theta, \phi)\) and \(\mathbf{a}_y(\theta, \phi)\) are the steering vectors along the \(x\)- and \(y\)-axes, respectively, and are given by:
\begin{equation}
\mathbf{a}_x(\theta, \phi) = \begin{bmatrix} 
1, & e^{j k_f d_x \sin\theta \cos\phi}, & \dots, & e^{j k_f d_x (\sqrt{M} - 1) \sin\theta \cos\phi}
\end{bmatrix}^T
\end{equation}
\begin{equation}
\mathbf{a}_y(\theta, \phi) = \begin{bmatrix} 
1, & e^{j k_f d_y \sin\theta \sin\phi}, & \dots, & e^{j k_f d_y (\sqrt{M} - 1) \sin\theta \sin\phi}
\end{bmatrix}^T.
\end{equation}

To construct the Jacobian matrix, we compute the partial derivatives of the steering vector \(\mathbf{a}(\theta, \phi)\) with respect to \(\theta\) and \(\phi\).

The partial derivative of \(\mathbf{a}_x(\theta, \phi)\) with respect to \(\theta\) is:
\begin{equation}
\begin{aligned}
\frac{\partial \mathbf{a}_x(\theta, \phi)}{\partial \theta} &= j k_f d_x \cos\theta \cos\phi \cdot \\
&\begin{bmatrix} 
0, & 1, & 2, & \dots, & \sqrt{M} - 1
\end{bmatrix} \odot \mathbf{a}_x(\theta, \phi),
\end{aligned}
\end{equation}
where \(\odot\) represents the element-wise product.

Similarly, the derivative of \(\mathbf{a}_y(\theta, \phi)\) with respect to \(\theta\) is:
\begin{equation}
\begin{aligned}
\frac{\partial \mathbf{a}_y(\theta, \phi)}{\partial \theta} &= j k_f d_y \cos\theta \sin\phi \cdot \\
&\begin{bmatrix} 
0, & 1, & 2, & \dots, & \sqrt{M} - 1
\end{bmatrix} \odot \mathbf{a}_y(\theta, \phi).
\end{aligned}
\end{equation}

The partial derivative of \(\mathbf{a}_x(\theta, \phi)\) with respect to \(\phi\) is:
\begin{equation}
\begin{aligned}
\frac{\partial \mathbf{a}_x(\theta, \phi)}{\partial \phi} &= -j k_f d_x \sin\theta \sin\phi \cdot \\
&\begin{bmatrix} 
0, & 1, & 2, & \dots, & \sqrt{M} - 1
\end{bmatrix} \odot \mathbf{a}_x(\theta, \phi),
\end{aligned}
\end{equation}
and the derivative of \(\mathbf{a}_y(\theta, \phi)\) with respect to \(\phi\) is:
\begin{equation}
\begin{aligned}
\frac{\partial \mathbf{a}_y(\theta, \phi)}{\partial \phi} &= j k_f d_y \sin\theta \cos\phi \cdot \\
&\begin{bmatrix} 
0, & 1, & 2, & \dots, & \sqrt{M} - 1
\end{bmatrix} \odot \mathbf{a}_y(\theta, \phi).
\end{aligned}
\end{equation}

The Jacobian matrix \(\mathbf{J}_{\mathbf{a}} \in \mathbb{C}^{M \times 2}\) is constructed by stacking the partial derivatives of the steering vector \(\mathbf{a}(\theta, \phi)\) with respect to \(\theta\) and \(\phi\):
\begin{equation}
\mathbf{J}_{\mathbf{a}} = 
\begin{bmatrix} 
\frac{\partial \mathbf{a}(\theta, \phi)}{\partial \theta} & \frac{\partial \mathbf{a}(\theta, \phi)}{\partial \phi}
\end{bmatrix}.
\end{equation}

The individual entries of the Jacobian are derived as:
\begin{equation}
\frac{\partial \mathbf{a}(\theta, \phi)}{\partial \theta} = \mathbf{a}_y(\theta, \phi) \otimes \frac{\partial \mathbf{a}_x(\theta, \phi)}{\partial \theta} + \frac{\partial \mathbf{a}_y(\theta, \phi)}{\partial \theta} \otimes \mathbf{a}_x(\theta, \phi),
\end{equation}

\begin{equation}
\frac{\partial \mathbf{a}(\theta, \phi)}{\partial \phi} = \mathbf{a}_y(\theta, \phi) \otimes \frac{\partial \mathbf{a}_x(\theta, \phi)}{\partial \phi} + \frac{\partial \mathbf{a}_y(\theta, \phi)}{\partial \phi} \otimes \mathbf{a}_x(\theta, \phi).
\end{equation}

This Jacobian matrix captures the sensitivity of the steering vector \(\mathbf{a}(\theta, \phi)\) to angular variations and plays a pivotal role in optimizing UAV positions in 3D space.

\end{appendices}

{\footnotesize
\bibliographystyle{IEEEtran}
\def\baselinestretch{0.9}
\bibliography{main}}

\end{document}